\documentclass[10pt]{IEEEtran}
\IEEEoverridecommandlockouts
\usepackage{cite}
\usepackage{amsmath,amssymb,amsfonts}
\usepackage{graphicx} 
\usepackage{textcomp}
\usepackage{amsmath}
\usepackage{booktabs} 
\usepackage[skip=1pt, font=footnotesize]{caption}
\usepackage{multirow}
\usepackage{hyperref}       
\usepackage{url}            
\usepackage{booktabs}       
\usepackage{amsfonts}       
\usepackage{nicefrac}       
\usepackage{microtype}      
\usepackage{xcolor}         
\usepackage{algorithm}
\usepackage{algpseudocode}
\usepackage{amsmath}
\usepackage{tikz}
\usepackage{quantikz}

\usepackage{orcidlink}
\usepackage{comment}
\usepackage{amssymb}
\usepackage{braket}
\usepackage{stfloats}
\usepackage{xcolor}
\def\BibTeX{{\rm B\kern-.05em{\sc i\kern-.025em b}\kern-.08em
    T\kern-.1667em\lower.7ex\hbox{E}\kern-.125emX}}

\usepackage{enumitem}
\setlist[itemize]{leftmargin=*}%
\setlist[enumerate]{leftmargin=*}%

\usepackage[compact]{titlesec}
\usepackage{stfloats}

\titlespacing\section{0pt}{0.2\baselineskip}{0.12\baselineskip}
\titlespacing\subsection{0pt}{0.15\baselineskip}{0.08\baselineskip}
\titlespacing\subsubsection{0pt}{0.1\baselineskip}{0.08\baselineskip}

\begin{document}
\pagestyle{empty} 
\bstctlcite{bstctl:etal, bstctl:nodash, bstctl:simpurl}

\title{PN-QNN: Harnessing Physical Noise as a Native Regularizer in Photonic Hybrid Quantum Neural Networks}

\author{\IEEEauthorblockN{Farah Elnakhal\IEEEauthorrefmark{1}\IEEEauthorrefmark{2}\orcidlink{0009-0006-5156-7324}, Alberto Marchisio\IEEEauthorrefmark{2}\IEEEauthorrefmark{3}\orcidlink{0000-0002-0689-4776}, Nouhaila Innan\IEEEauthorrefmark{2}\IEEEauthorrefmark{3}\orcidlink{0000-0002-1014-3457}, Gabriel Falcao\IEEEauthorrefmark{1}\IEEEauthorrefmark{4}\orcidlink{0000-0001-9805-6747},
Muhammad Shafique\IEEEauthorrefmark{2}\IEEEauthorrefmark{3}\orcidlink{0000-0002-2607-8135}}

\IEEEauthorblockA{\IEEEauthorrefmark{1} \normalsize Science Division, New York University Abu Dhabi, UAE\\
\IEEEauthorblockA{\IEEEauthorrefmark{2} \normalsize eBrain Lab, Division of Engineering, New York University Abu Dhabi, PO Box 129188, Abu Dhabi, UAE\\}
\IEEEauthorblockA{\IEEEauthorrefmark{3} \normalsize Center for Quantum and Topological Systems, NYUAD Research
Institute, New York University Abu Dhabi, UAE\\}
\IEEEauthorblockA{\IEEEauthorrefmark{4} \normalsize Instituto de Telecomunicações, University of Coimbra, Portugal\\
Emails: \{fqe9080, alberto.marchisio, nouhaila.innan, gabriel.falcao, muhammad.shafique\}@nyu.edu}}
\vspace{-25pt}
}

\maketitle
\thispagestyle{empty}

\begin{abstract}
Physical noise in near-term quantum hardware is usually treated as a nuisance to suppress. We ask whether it can instead act as a hardware-native regularizer for photonic hybrid quantum-classical neural networks (PHQCNNs), analogous to noise-injection regularization in classical deep learning. Using Quandela's Perceval simulator and the MerLin framework, we build PHQCNNs for Iris, Digits, and MNIST and inject Perceval's seven-parameter physical noise model directly into training. A genetic algorithm searches the six continuous noise dimensions and 1 boolean parameter to find, per dataset, the configuration maximizing validation accuracy, compared against a noiseless baseline across five seeds. GA-tuned noise yields modest accuracy gains on Iris (+0.82pp) and Digits (+1.45pp), but a clear degradation on MNIST ($-$1.21pp). Per-parameter sweeps show that no individual noise parameter is consistently beneficial, motivating the joint search, while a second-order loss expansion shows that physical noise induces a Tikhonov-like regularization term whose effect is dataset-dependent. Physical photonic noise can thus act as a free regularizer, but not universally.

\end{abstract}


\section{Introduction}

Noise is the central obstacle standing between today's noisy intermediate-scale quantum (NISQ) devices and quantum advantage: decoherence, photon loss, and gate imprecision are typically framed as sources of error to be suppressed, mitigated, or corrected away \cite{Preskill_2018}. In classical deep learning, however, the relationship between noise and model quality is far less one-sided. Injecting noise into the inputs, weights, or activations of a classical neural network during training is a well-established regularization technique, provably equivalent under mild assumptions to a smoothing (Tikhonov-like) penalty on the learned function \cite{bishop}, and an analogous dropout mechanism has recently been proposed for quantum neural networks by randomly removing gates or entangling connections during training \cite{Kobayashi_2022}. This raises a natural question for photonic hybrid quantum classical neural networks (PHQCNNs): \textbf{rather than treating the physical noise inherent to a linear-optical processor purely as a nuisance, can it instead be characterized and, within limits, exploited as a free, hardware-native regularizer?}

Our contributions are as follows: (1) \textit{we reframe Perceval's physical noise model}, typically used only to characterize hardware fidelity, \textit{as a tunable regularization mechanism for PHQCNNs, and inject it directly into training and evaluation}; (2) \textit{we design a genetic algorithm} that searches the joint noise-parameter space under a physically-motivated gene grouping (source, interferometer, global) \textit{to find per-dataset noise configurations maximizing validation accuracy}, evaluated on Iris, Digits, and MNIST; and (3) \textit{we provide both empirical evidence}, showing a small positive effect on Iris and Digits and a negative effect on MNIST, \textit{and a theoretical account}, via a second-order expansion of the training loss, \textit{of when and why physical noise should be expected to act as an implicit Tikhonov-like regularizer}.

\section{Background and Related Work}

\subsection{Classical noise injection as a regularizer} 

Training a neural network with noise added to its inputs is, to leading order, equivalent to adding an explicit smoothing penalty to the training objective, effectively trading a small amount of training-set fit for a flatter, more robustly generalizing solution \cite{bishop}. More recent work has extended this analysis to noise injected at arbitrary layers of a deep network and derived the explicit form of the induced regularizer \cite{camuto2021explicitregularisationgaussiannoise}, and has proposed structured noise-injection schemes with provable robustness benefits \cite{JMLR:v15:srivastava14a}. We treat this literature as the classical analogue of the phenomenon we study here.

\subsection{Quantum dropout} 

Within quantum machine learning, the closest classical-inspired regularization concept is quantum dropout, in which gates, qubits, or entangling connections are randomly removed during training to limit the expressibility of a parameterized quantum circuit and thereby curb overfitting~\cite{scala2025general}. While quantum dropout changes the structure of the circuit being trained, the noise as a regularizer mechanism we study here uses noise to perturb how faithfully that fixed circuit is physically realized (photon loss, partial distinguishability, phase drift). We view the two as complementary rather than competing mechanisms.
 
\subsection{Photonic Hybrid quantum neural networks and noise modeling} 

\begin{figure}[htpb]
\centering
\includegraphics[width=0.95\columnwidth]{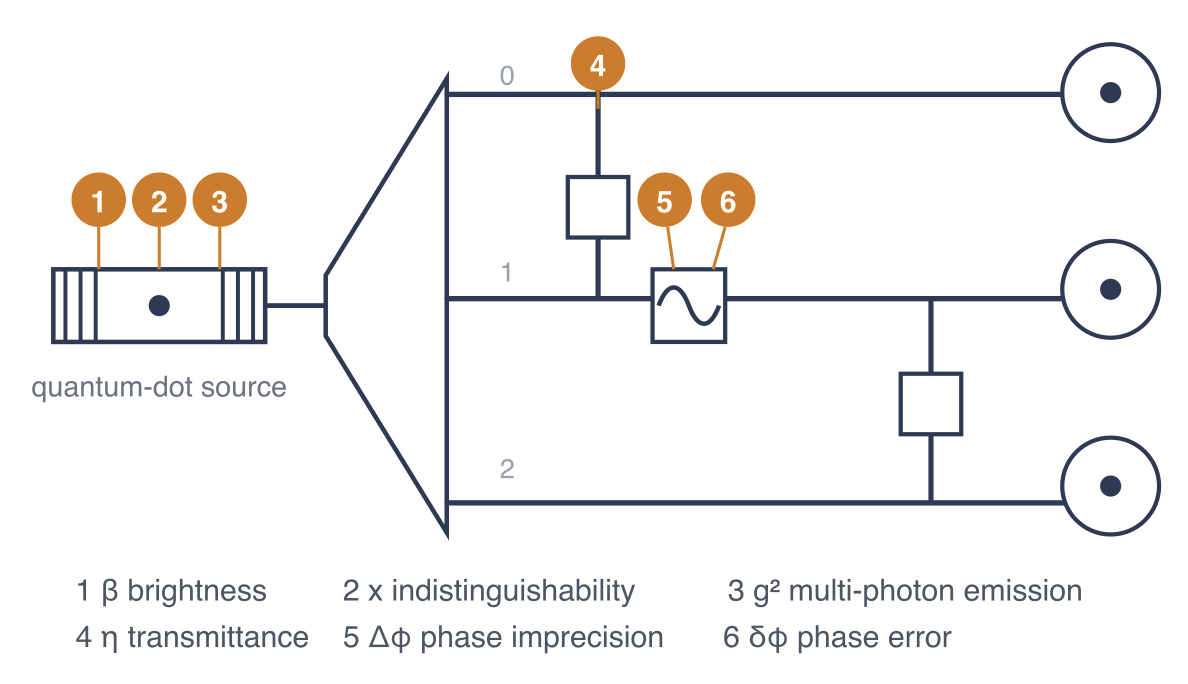}
\caption{Noise model on photonic GBS device. The Perceval noise model contains 7 noise parameters (6 continuous values and 1 boolean value), which can either be located at the photon source, interferometer, or globally.}
\label{fig:noise_model}
\end{figure}

Our PHQCNN architectures are implemented using Quandela's Perceval simulation library~\cite{heurtel2023perceval} together with the MerLin framework for differentiable, PyTorch-native photonic quantum layers~\cite{notton2026merlin}. The seven-parameter physical noise model we use (brightness, indistinguishability, $g^{(2)}$, $g^{(2)}$ distinguishability, transmittance, phase imprecision, phase error) (see Fig~\ref{fig:noise_model}) follows the standard characterization of near-term linear-optical hardware imperfections~\cite{somaschi2016near} used to assess photonic device quality and to simulate noisy linear-optical circuits more generally \cite{heurtel2023perceval, maring2023generalpurposesinglephotonbasedquantumcomputing}.

\section{Methodology}

\subsection{Datasets and Architectures}
 
All models are built on Quandela's linear optical stack (Perceval~\cite{heurtel2023perceval} and MerLin~\cite{notton2026merlin}). Each architecture consists of a \texttt{QuantumLayer} implementing a photonic circuit with a trainable entangling layer, angle encoding of the input features onto the optical modes, and a second entangling layer. Therefore, the input is encoded between two trainable entangling layers operating on Fock states.

We use three different architectures, one for each dataset (Iris~\cite{iris_53}, Digits~\cite{uci_digits_1998}, and MNIST~\cite{lecun-mnisthandwrittendigit-2010}); see Table~\ref{tab:datasets}.
 
For Iris, four standardized features are angle-encoded directly into 6 modes with a 3-photon Fock input state. The \texttt{ML.QuantumLayer} \cite{notton2026merlin} output passes through a grouping and readout layer to produce 3 class logits, without any additional classical pre- or post-processing.

For Digits, the 64-dimensional inputs are reduced to 8 principal components via PCA (fit on the training split). A classical feature extractor, composed of \texttt{BatchNorm1d}~\cite{ioffe2015batchnormalizationacceleratingdeep} and MLP (\texttt{Linear $\rightarrow$ SiLU $\rightarrow$ Linear})~\cite{elfwing2017sigmoidweightedlinearunitsneural}, remaps features before encoding into an 8-mode, 4-photon \texttt{QuantumLayer}, followed by a classical head (\texttt{Linear $\rightarrow$ GELU $\rightarrow$ Dropout(0.1) $\rightarrow$ Linear($\rightarrow$10)}) \cite{JMLR:v15:srivastava14a} \cite{hendrycks2023gaussianerrorlinearunits}.

For MNIST, raw pixels are flattened, standardized, and PCA-reduced to 16 components. A classical MLP (\texttt{BatchNorm1d} $\rightarrow$ two \texttt{Linear$\rightarrow$GELU} blocks with dropout) maps the features to 8 values, which are converted into phases via a learnable per-mode encoding \texttt{sigmoid}$(x \cdot s + b)\cdot\pi$ \cite{elfwing2017sigmoidweightedlinearunitsneural}. These phases are processed by an 8-mode \texttt{QuantumLayer}, followed by a classical head (\texttt{Linear $\rightarrow$ GELU $\rightarrow$ BatchNorm1d $\rightarrow$  Linear($\rightarrow$10)}).

\subsection{Noise Model and Injection}
 
Noise is injected by attaching a \texttt{pcvl.NoiseModel} \cite{heurtel2023perceval} to the quantum layer's experiment object after the layer is constructed (before training). The noise model contains the seven physical parameters used by Perceval, which are as follows:

\begin{itemize}
    \item Brightness: loss at the first step in photon generation.
    \item Indistinguishability: chance that 2 photons are indistinguishable.
    \item $g^{(2)}$: Second order intensity autocorrelation at 0 time delay; how often 2 photons are emitted by the source instead of 1 
    \item $g^{(2)}$ distinguishable: if generated photon is distinguishable
    
    \item Transmittance: global loss applied across the whole system
    \item Phase imprecision: maximum precision of the phase shifters due to calibration imperfection
    \item Phase error: maximum random noise on the phase shifters, modeling thermal drift and electronic jitter in heaters.
\end{itemize}
 
Once the noise model is attached, the same model instance, under the same injected noise, is active during both training and evaluation forward passes, and is used to compute both train and test accuracy.
 
\subsection{Per-parameter sensitivity sweep}
 
For each of the seven noise parameters, 20 values are swept linearly from [0, 0.95] in steps of 0.05, with a fresh model trained from the same initialization for each value, swept one parameter at a time rather than jointly for both Iris and Digits at 100 training epochs (Adam \cite{kingma2017adammethodstochasticoptimization}, lr=0.005 (Iris), 0.008 (Digits), tuned per dataset).
 
\subsection{Noise-parameter optimization}
 
A genetic algorithm (GA) \cite{1975-26618-000} searches the joint 7-dimensional noise parameter space, with 6 continuous variables and a binary toggle for $g^{(2)}$ distinguishability, to find the single configuration that maximizes the validation accuracy for each dataset, using a train/validation/test split.
 
Genes are grouped by physical origin: source (\texttt{brightness}, \texttt{indistinguishability}, \texttt{$g^{(2)}$}), global (\texttt{transmittance}) and interferometer (\texttt{phase\_imprecision}, \texttt{phase\_error}). Since this grouping is what structures the crossover, recombination respects the physical subsystem boundaries, rather than mixing genes uniformly. Selection uses tournament selection (\texttt{k=3}) \cite{Miller1995GeneticAT, GOLDBERG199169} with elitism (top 2 individuals retained each generation) and crossover ($ P(C) = 0.8$). We employ Gaussian mutation \cite{489178} with mutation rate $p_m=0.4$ and mutation strength $\sigma$, where $\sigma$ decays by a factor of $0.95$ each generation, together with niching~\cite{10.1162/evco.2008.16.3.315} to preserve population diversity and mitigate premature convergence.
 
For all datasets, the GA uses a population of 20 evolved over 25 generations, with each candidate trained for 100 epochs. The selected configuration is then retrained from scratch using the same epoch budget and circuit architecture. Each found best noise configuration is retrained from scratch and evaluated on the held-out test split, and compared against a noiseless model evaluated identically. 

\section{Results and Discussion}
\subsection{Experimental Setup}

\begin{table}[t]
\centering
\caption{Dataset preprocessing, split and encoding.}
\label{tab:datasets}
\resizebox{\columnwidth}{!}{%
\begin{tabular}{@{}llccl@{}}
\toprule
Dataset & Preprocessing & Classes & Train/Val/Test Split & Modes/Photons \\
\midrule
Iris   & 4 features, standardized (train)            & 3  & 60:20:20 & 6 modes, 3 photons \\
Digits & 64px $\rightarrow$ SS $\rightarrow$ PCA(8), fit on train  & 10 & 60:20:20 & 8 modes, 4 photons \\
MNIST  & 784px $\rightarrow$ SS $\rightarrow$ PCA(16), fit on train & 10 & 75:15:10 & 8 modes, 8 photons\\
\bottomrule
\end{tabular}%
}
\end{table}

\subsubsection{Training Protocol}

Adam \cite{kingma2017adammethodstochasticoptimization} is used as an optimizer throughout, with a small weight-decay term ($1\mathrm{e}{-4}$) applied uniformly across all conditions, including the noiseless case. Consequently, the noiseless baseline already carries a modest classical $L_2$ regularizer, and the noise-injected conditions are compared against a regularized rather than an unregularized baseline. 

All reported conditions are run at 5 seeds each, with final accuracy averaged. Experiments run in Perceval's simulator backend under exact permanent-based (SLOS) \cite{Heurtel_2023} computation rather than sampling, so reported probabilities are simulator exact given the specified noise model, not subject to additional shot noise from a finite sampling budget.

\subsubsection{Reporting}

For each dataset and condition, we report final test accuracy, and qualitatively describe the train/test accuracy gap from training curves as a regularization signature. Per parameter sensitivity sweeps are reported as accuracy versus parameter-value curves.

\subsection{Main Quantitative Results}

Table~\ref{tab:main-results} summarizes final accuracy across conditions; GA-tuned noise helps on Iris and Digits but hurts on MNIST. We observe a 0.82pp accuracy increase in Iris, 1.45 pp in Digits, but a decrease of 1.21pp in MNIST.

\begin{table}[t]
\centering
\caption{Main quantitative results across datasets. Epoch budget is matched within each dataset (noiseless vs.\ GA-best noise)}
\label{tab:main-results}
\resizebox{.6\columnwidth}{!}{%
\begin{tabular}{@{}llccc@{}}
\toprule
Dataset & Condition & Final Test Accuracy \\
\midrule
\multirow{2}{*}{Iris}
 & Noiseless      & 95.54\% $\pm$ 0.38\%\\
 & GA-best noise  & 96.36\%  $\pm$ 0.26\%\\
\midrule
\multirow{2}{*}{Digits}
 & Noiseless      & 94.71\% $\pm$ 0.16\%\\
 & GA-best noise  & 96.16\% $\pm$ 0.19\%\\
\midrule
\multirow{2}{*}{MNIST}
 & Noiseless      & 97.10\% $\pm$ 0.61\%\\
 & GA-best noise  & 95.89\% $\pm$ 0.27\%\\
\bottomrule
\end{tabular}
}
\end{table}

\subsection{Noise Configuration Found by the GA}

Table~\ref{tab:ga-configs} shows the GA-found noise configuration used for the GA-best noise results above, for each dataset.

\begin{table}[t]
\centering
\caption{GA-found noise configuration by dataset. All parameters range [0,1] except phase error [0, $\pi$] and $g^{(2)}$ distinguishable (boolean)}
\label{tab:ga-configs}
\resizebox{\columnwidth}{!}{%
\begin{tabular}{@{}lccccccc@{}}
\toprule
Dataset & Bright. & Indist. & $g^{(2)}$ & $g^{(2)}$ dist. & Transmit. & Ph.\ Imprec. & Ph.\ Error \\
\midrule
Iris   & 0.776 & 0.265 & 0.473 & 1 & 0.616 & 0.902 & 1.585 \\
Digits & 0.309 & 1.000 & 0.860 & 1 & 0.480 & 0.0732 & 1.081 \\
MNIST  & 0.716 & 0.886 & 0.310 & 1 & 0.730 & 0.905 & 2.148 \\
\bottomrule
\end{tabular}%
}
\end{table}

The three configurations are structurally different from one another: brightness ranges from 0.31 to 0.78 across the three datasets, phase imprecision from 0.07 to 0.90, and phase error from 1.08 to 2.15. No single parameter setting is shared across datasets, which is consistent with the noise regularization effect (or lack thereof) being dataset- and architecture-dependent rather than governed by one universally good noise profile.

\subsection{Per-Parameter Sensitivity Sweep}

\begin{figure*}[t!]
\centering
\includegraphics[width=\linewidth]{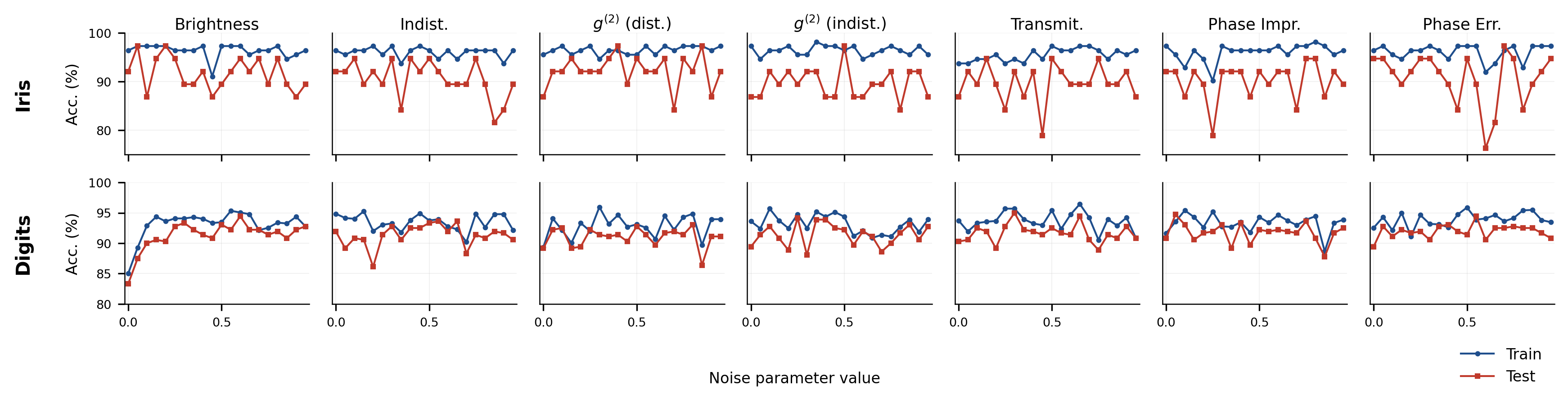}
\caption{Single parameter sensitivity sweep on Iris (top) and Digits (bottom). All noise parameters are fixed at their noiseless values except the parameter being swept, and accuracy is recorded.}
\label{fig:sweep}
\end{figure*}

Figure~\ref{fig:sweep} shows accuracy as a function of each of the seven noise parameters in isolation, swept over $[0, 0.95]$ with all other parameters held at their noiseless value, for Iris (top) and Digits (bottom). No single parameter shows a monotonic, uniformly-beneficial trend across its full range for either dataset: accuracy degrades or plateaus for some parameters and stays flat for others, and the per-parameter picture on its own does not predict which joint configurations the GA subsequently found to be strong. This is the empirical basis for searching all seven dimensions jointly rather than tuning parameters independently.

\subsection{Training Dynamics by Condition}

\begin{figure*}[htbp]
\centering
\includegraphics[width=\linewidth]{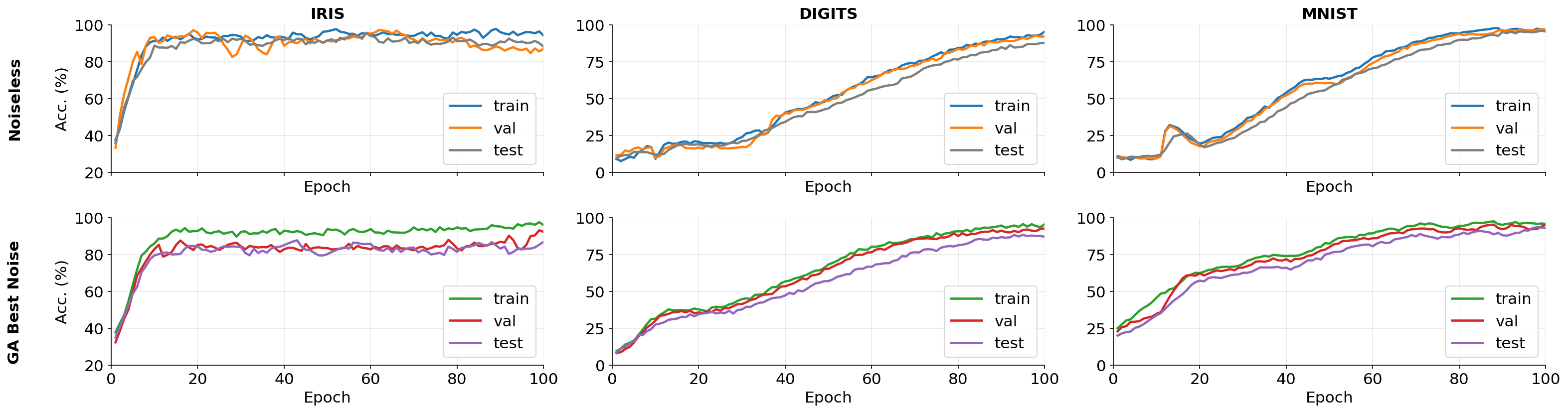}
\caption{Left: Iris training curves, noiseless (top) vs.\ GA-best noise (bottom). Middle: Digits training curves, noiseless (top) vs.\ GA-best noise (bottom). Right: MNIST training curves, noiseless (top) vs.\ GA-best noise (bottom). All trained for 100 epochs.}
\label{fig:iris-conditions}
\end{figure*}

Figure~\ref{fig:iris-conditions} shows accuracy (train and test) for the noiseless (top) and GA-best noise (bottom) conditions, for all datasets.

For Iris, both conditions reach comparable train accuracy by the end of training, while test accuracy is noisier and somewhat lower under GA-best noise for much of the run. For Digits, the GA-best noise condition visibly leads the noiseless baseline through the middle portion of training before the two converge to a similar range by the final epochs, consistent with the modest final-accuracy edge in Table~\ref{tab:main-results}.

Unlike Iris and Digits, MNIST's noiseless and GA-best trajectories are not superimposable: the GA-best run trains more slowly through the early epochs and settles to a visibly lower final accuracy under the same 100-epoch budget.

\subsection{Discussion}
\label{sec:theory}

\subsubsection{Why physical noise can regularize}

Writing the trainable circuit as a mode-unitary $U(\theta) \in U(m)$ acting on $m$ modes with fixed input Fock state $S=(s_1,\dots,s_m)$, the noiseless probability of an output $T=(t_1,\dots,t_m)$ is given by a matrix permanent \cite{aaronson2010computationalcomplexitylinearoptics},
\begin{equation}
P(T\mid S,U) = \frac{|\mathrm{Perm}(U_{S,T})|^2}{s_1!\cdots s_m!\,t_1!\cdots t_m!},
\label{eq:permanent}
\end{equation}
and training minimizes $\mathcal{L}(\theta)=\mathbb{E}_{(x,y)\sim\mathcal{D}}[\ell(f_\theta(x),y)]$ over this noiseless model. The seven physical noise parameters perturb what is actually realized per run without changing $U(\theta)$'s functional form: brightness/transmittance attenuate photon number (equivalent to coupling each mode to a traced-out environment mode); indistinguishability and $g^{(2)}$ interpolate between fully-coherent multiphoton interference and a partially distinguishable-photon sampling distribution; phase imprecision/error perturb the entries of $U(\theta)$ by a small stochastic offset each run. Schematically, $f_\theta^{\text{noise}}(x) = f_\theta(x) + \xi(\theta,x)$ for an architecture- and parameter-dependent, non-Gaussian perturbation $\xi$. A second-order expansion of the expected loss under this perturbation gives, to leading order, $\mathbb{E}_\xi[\ell(f_\theta(x)+\xi,y)]
\approx\;
\ell(f_\theta(x),y)+\nabla_f \ell(f_\theta(x),y)^\top\mu_\xi(\theta,x) 
+\tfrac12
\mathrm{tr}\!\left(\Sigma_\xi(\theta,x)\nabla_f^2 \ell(f_\theta(x),y)\right),$
where $\mu_\xi(\theta,x)=\mathbb{E}[\xi]$ and
$\Sigma_\xi(\theta,x)=\mathbb{E}[(\xi-\mu_\xi)(\xi-\mu_\xi)^\top]$.
When $\mu_\xi(\theta,x)\approx 0$, the first-order term vanishes, leaving a hardware-induced regularization effect whose strength depends jointly on the seven physical-noise parameters, the input $x$, the architecture, and $\theta$ in a non-separable way.

\subsubsection{What the data actually supports}

Three findings matter most:

\begin{enumerate}
\item Iris shows a small, real benefit (+0.82pp).
\item Digits shows a small, real benefit (+1.45pp).
\item MNIST shows a reliable, negative effect (-1.21pp).
\end{enumerate}

Altogether, this is evidence for a modest, positive regularization effect on Iris and Digits and a negative effect on MNIST, with the GA finding a structurally different noise configuration for each dataset (Table~\ref{tab:ga-configs}) rather than one setting that generalizes across problems, suggesting that the noise-induced regularization effect is dataset- and architecture-dependent rather than a general property of physical noise.

\subsubsection{Limitations}

Two main limitations apply: (1) the three architectures differ in mode count, photon number, and classical head depth, so the difference between Iris/Digits and MNIST is confounded with architecture rather than isolated to dataset noise-sensitivity alone; and (2) this work is simulation based; as of now, noise perturbations on hardware are not physically available.

\section{Conclusion}

We investigated whether the physical noise inherent to photonic quantum hardware can be characterized and harnessed as a native regularizer for photonic hybrid quantum-classical neural networks. Using Perceval's seven-parameter noise model and a genetic algorithm to search the six continuous noise dimensions and one boolean parameter, we found dataset-dependent effects: GA-tuned noise configurations produced small but consistent accuracy gains on Iris and Digits, while producing a clear accuracy loss on MNIST under matched training budgets. Per-parameter sensitivity sweeps showed no individual noise parameter is uniformly beneficial, and the GA converged to structurally distinct configurations for each dataset rather than a single noise profile that transfers across problems. These results indicate physical noise can act as a free, hardware-native regularizer, but its benefit is not guaranteed and appears to depend on the interaction between the noise channel, dataset, and circuit architecture. Future work should extend the joint search to larger and more diverse datasets and circuit sizes, incorporate additional reporting to more directly quantify the regularization effect, and examine whether the theoretical second-order expansion in Section~\ref{sec:theory} can predict, rather than only explain, which datasets will benefit from a given noise profile.

\section*{Acknowledgment}
This work was supported in part by the NYUAD Center for Interdisciplinary Data Science \& AI (CIDSAI), funded by Tamkeen under the NYUAD Research Institute Award CG016, by the NYUAD Center for Quantum and Topological Systems (CQTS), funded by Tamkeen under the NYUAD Research Institute grant CG008, and by the NYUAD Center for CyberSecurity (CCS), funded by Tamkeen under the NYUAD Research Institute Award G1104. This work is partially funded by national funds through FCT – Fundação para a Ciência e a Tecnologia, I.P., and, when eligible, co-funded by EU funds under project/support UID/50008/2025 – Instituto de Telecomunicações, with DOI identifier - https://doi.org/10.54499/UID/50008/2025.

\bibliographystyle{IEEEtran}

\bibliography{refs}

\end{document}